# Edge Contact Forces and Quasi-Balanced Power

F. DELL'ISOLA[1] and P. SEPPECHER[2]
[1]*Università di Roma La Sapienza, Dipartimento di Ingegneria Strutturale e Geotecnica, Via Eudossiana 18; 00184 Roma, Italy*
[2]*Université de Toulon et du Var, Laboratoire d'Analyse Non Linéaire Appliquée; 83957 La Garde Cedex, France*

**Abstract.** We consider continuous media in which contact edge forces are present. Introducing the notion of quasi-balanced contact force distribution, we are able to prove the conjectures by Noll and Virga [1] concerning the representation of contact edge forces. We generalize the Hamel–Noll theorem on the Cauchy postulate. Then we adapt the celebrated tetrahedron construction of Cauchy in order to obtain a representation theorem for stress states. In fact, we show that two stress tensors of order two and three are necessary for such a representation. Moreover we f nd the relationship between the notion of interstitial working introduced by Dunn and Serrin [2] and the notion of contact edge force.

**Sommario.** Si considerano continui in cui sono presenti forze di contatto di spigolo. Una volta introdotta la nozione di distribuzione di forza di contatto quasi-bilanciata, diventa possibile la dimostrazione delle congetture avanzate da Noll e Virga [1] riguardo la rappresentazione delle forze di spigolo. I nostri ragionamenti si basano su una generalizzazione del teorema di Noll sul postulato di Cauchy e su una rielaborazione della celebre costruzione del tetraedro di Cauchy. In tal modo viene dimostrato un teorema di rappresentazione degli stati di tensione in cui appaiono due tensori uno del secondo, l'altro del terzo ordine. Come corollario di tale teorema di rappresentazione si determina la relazione fra la nozione di lavoro interstiziale introdotta da Dunn e Serrin [2] e la forza di contatto di spigolo.

## 1. Introduction

A crucial concept in continuum mechanics is that of stress. Engineers working in strength of materials have used this concept so frequently that it has become as intuitive as the notion of force. However, its introduction was not trivial, being based on some fundamental assumptions whose weakening raises interesting problems. In this paper we address the case in which contact force interactions are represented not only by a surface density, but also by a line density define on the edges of the contact surface (if any).

Stress tensor and contact force are interrelated notions. Cauchy postulated (cf. [3], [5]) that the contact force exerted on a body can be represented by a surface density of force localized on the contact surface depending only on the normal (and the position). Then he showed, using a family of tetrahedrons with vanishing size, that balance of force implies a linear dependence of the surface density of contact force on the normal to the contact surface. The stress tensor is a representation of this linear function.

Noll [6] proved the Cauchy postulate assuming that

(i) the contact force is absolutely continuous with respect to the Hausdorff bidimensional measure define on Cauchy cuts;
(ii) there is an upper bound for the norm of the surface density of contact force when it is regarded as a function define in the set of the shapes of Cauchy cuts.

In this paper we propose a proof of Cauchy postulate (Theorem 4) valid without the a priori assumption that the contact force density is bounded independently of the shape of the contact surface. Our weaker assumptions – in particular – allow for the linear dependence of surface density of contact forces on the local mean curvature of Cauchy cuts. In [1] Noll and Virga announced this generalization, citing a forthcoming paper which we could not fin in the literature.

The results of Cauchy and Noll strongly depend on the implicit assumption of vanishing contact edge forces. Noll and Virga in [1] developed a theory in which general edge contact interactions are considered. Here we limit ourselves to interactions between a body and its exterior (for a discussion of the more general case see [7]). Moreover we only consider relatively regular contact surfaces (see Section 2).

Noll and Virga in their paper [1] declare: 'It would be desirable to derive [Assumptions I to IV] from other assumptions that have more transparent and natural physical interpretations'. Here we face this diff culty by considering the power expended by contact forces. The theorem of kinetic energy states that the power of contact forces is balanced by the sum of the stress power, the power of inertial forces and the power of external body forces [3]. It is physically reasonable to assume that these three quantities are volume-continuous. Therefore, we assume that the power $P_U^c$ expended by the distribution of contact force on any $C^\infty$ velocity fi ld $U$ is quasi-balanced: then there exists a scalar $K_U$ such that, for any admissible domain (see Defi ition 0 in Section 2) $V$, the inequality $P_U^c(V) < K_U |V|$ holds, $|V|$ denoting the volume of $V$.

The problem is now to determine a correct expression for the stress power. We fi st show that it cannot simply equal the sum of the powers of surface and edge force. Then we show that the edge force can only be in conjunction with a supplementary supply of mechanical energy. This supply corresponds to the 'interstitial working' introduced by Dunn and Serrin [2], [8]. In the literature it is recognised that a proper modelling of some media requires the introduction of such a supply: this is the case, for instance, of Cahn–Hilliard fluid [9], [10]. However, the hypothesis made by Dunn and Serrin [2] that this energy supply can be represented as a flu is nowhere justif ed, to our knowledge. As far as we know, the close relationship between interstitial working and edge force was not yet pointed out. In this paper, we prove that nonzero contact edge forces require an interstitial energy f ux; we also give the most general form of interstitial working compatible with zero edge forces.

Instead of introducing the notion of interstitial working from the beginning, we prefer to describe contact interactions in terms of force distributions which are not necessarily measures. This approach, which is close to the one by Dunn and Serrin, seems to have the following advantages:

(i) we do not assume a priori that the supplementary energy supply is a flux
(ii) the linear dependence of the supplementary energy supply on the velocity fie d is for us a basic assumption (for a discussion about this dependence, refer to [8]), so that the mechanical nature of the interstitial working is clearly understood;
(iii) the nature of boundary conditions gets clearer [11] [12].

Contact force distributions may be interpreted as an asymptotic limit of nonlocal short-range forces. In general this limit is a distribution (in the sense of Schwartz), whose support is contained in the contact surface and whose order may be greater than zero. We give a trivial example of such an asymptotic analysis in sub-section 4.1. Note that the introduction of force distributions of order greater than zero is not new in continuum mechanics: e.g., the brothers

Cosserat introduced [13] a surface density of couples, which is a special type of fi st order surface force distribution.

From a mathematical point of view, the main assumption of this paper can be stated as follows: the zero-order part (with respect to the normal derivative) of the contact force distribution is the sum of two measures absolutely continuous with respect to $\mathcal{H}^2(S^r)$ and $\mathcal{H}^1(L^r)$, respectively, where $S^r$ is the regular part of the contact surface, $L^r$ is the regular part of contact edges (and $\mathcal{H}^k$ denotes the $k$-dimensional Hausdorff measure). Indeed, a decomposition theorem for distributions whose support is a $C^\infty$ manifold [14] uniquely determines the zero-order part (with respect to normal derivatives) of a contact distribution.

Section 3 is devoted to the proof that such a sum of measures cannot be a quasi-balanced distribution unless the (line) measure on the edges vanishes.

In Section 4 we add to the above define sum of zero-order measures a distribution which is of the order of one in the normal derivative and call it a normal distribution. Such a term was f rst introduced by Germain, who called it a surface (normal) double traction on page 557 of [16]. Under the assumption of quasi-balance of the contact force distribution we f nd the relation between surface normal double traction and contact edge force. We show that:

(i) the normal distribution depends only on the normal to the surface (Theorem 6),
(ii) there exists a three-tensor fi ld $C$ (a hyper-stress tensor) which represents the normal distribution (Theorem 7),
(iii) the edge force can be expressed in terms of the tensor $C$, and depends on the shape of the contact wedge only through the dihedron tangent to it (Theorem 8),
(iv) there exists a two-tensor fie d $T$ (ordinary stress tensor) which determines that part of the surface contact force which depends linearly on the normal to the contact surface (Theorem 9), the remaining part being determined by $C$.

## 2. Hypotheses and Notations

DEFINITION 0. *We call* domains ([3]) *the closures of open Kellogg [4] regular regions.*

We want to describe the contact force exerted on a subbody, identif ed with the domain it presently occupies, through its boundary $S$ (the contact surface). We are interested in their dependence on the shape of $S$. Then we have to defin precisely what we call shape of $S$ at a point $x \in S$.

We say that the shape of the oriented surface $S$ at the point $x \in S$ is the same as the shape of $S'$ at the point $x' \in S'$ if and only if there exists a neighbourhood of $x'$ in which $t_{x'-x}(S)$ coincides (as an oriented surface) with $S'$ ($t_u$ denoting the translation by the vector $u$).

DEFINITION 1. *We call shape of $S$ at the point $x \in S$ the equivalence class of $(S, x)$ with respect to the above define relation. We denote it $(\widetilde{S, x})$. Note that, according to this defin tion, rotations change shape.*

DEFINITION 2. *Plane shape: The shape of a plane $P$ at any of its points depends only on the normal $n$ to the plane. When there is no ambiguity, we denote it simply by $n$.*

DEFINITION 3. *Dihedral shape: Let us consider a non-degenerate dihedron. We denote by $n_1$ and $n_2$ the external normals to the half-planes forming it, and by $\nu_1$ and $\nu_2$ the external normals to the boundary of these half-planes within their planes. We denote by $\tau$ the unit* *vector tangent to the edge, such that* $\nu_1 = \tau \times n_1$ *and* $\nu_2 = -\tau \times n_2$. *On the edge of such a dihedron, the shape is constant and is determined by* $n_1$, $n_2$ *and* $\tau$. *This shape will be denoted by* $(n_1, n_2, \tau)$. *Note that* $(n_1, n_2, \tau) = (n_2, n_1, -\tau)$. *The angle* $(-n_1, n_2)$ *in the plane oriented by* $\tau$ *will be called the dihedral angle of* $(n_1, n_2, \tau)$; *it is different from* 0, $\pi$ *or* $2\pi$.

DEFINITION 4. *Cuts: Let $V$ and $V'$ be two domains whose boundaries are the surfaces $S$ and $S'$. Let $S''$ be the boundary of $V \cap V'$. At each point $x$ in $S \cap S' \cap S''$ the shape of $S''$ depends only on the shape of $S$ and $S'$. We denote the shape of $S''$ by $Cut((\widetilde{S,x}),(\widetilde{S',x}))$.*

*Remark.* We will only use cuts of $S$ with surfaces whose shape is a plane shape $u$. We call them *plane cuts* and write by $Cut((\widetilde{S,x}), u)$. In the following we will always deal with surfaces which are boundaries of Kellogg regular regions, also when these surfaces are obtained by means of cuttings.

DEFINITION 5. *Admissible domains: We only consider domains whose boundary $S$ (admissible surface) is a finite union of two-dimensional $C^\infty$ compact manifolds with boundary (called the faces of $S$) and such that the union of the boundaries of these faces is a finite union of one-dimensional $C^\infty$ compact manifolds with boundary (called the edges of $S$). The set of all internal points of the faces (regular points of the surface, or points where $S$ has a regular shape) is denoted by $S^r$; the set of all internal points of the edges (regular points of the edges, or points where $S$ has an edge shape) is denoted by $L^r$. Moreover, we assume that, everywhere in $L^r$, $S$ is tangent to a non degenerate dihedron.*

DEFINITION 6. *Sets of shapes: Let $S$ be an admissible surface. We denote by $\Phi^r(S)$ the set of all regular shapes of $S$ and by $\Phi^e(S)$ the set of all edge shapes of $S$*

$$\Phi^r(S) := \{(\widetilde{S,x}); x \in S^r\}, \qquad \Phi^e(S) := \{(\widetilde{S,x}); x \in L^r\}.$$

As vertices are not involved in our reasoning we will call *set of shapes* of a surface $S$ the set $\Phi(S) = \Phi^r(S) \cup \Phi^e(S)$.

DEFINITION 7. *Prescribed shapes: A set of shapes $E$ is called a set of prescribed shapes if there exists a finite set $\{S_1, \ldots, S_m\}$ of admissible surfaces such that $E \subset \bigcup_{i=1}^m \Phi(S_i)$.*

EXAMPLES. The set of images $C_t$ of a cube $C_1$ under a family of homothetic transformations of ratio for $t \in ]0, 1]$ is a set of prescribed shapes; the set of images $C'_t$ of a cube $C_1$ under a family of rotations of angle $t \in [0, \pi]$ around a given axis is not a set of prescribed shapes; the family of spheres $S_t$ of centre $x_0$ and radius $t \in ]0, 1]$ have not prescribed shapes.

A set of shapes $E$ is called a set of *prescribed plane cuts* if there exist a finite sequence $(S_i)$ of admissible surfaces and a finite sequence $(u_j)$ of unit vectors such that

$$E \subset \bigcup_{i,j} \{Cut(f, u_j); f \in \Phi^r(S_i)\}.$$

*Remark.* In the following we consider surface contact forces depending continuously on curvature. We will assume that such forces densities are bounded on a set of prescribed shapes but not on the set of all shapes.

DEFINITION 8. *Contact force: The contact forces $F^c$ exerted on any admissible domain V through its boundary S will be determined by two vector functions, F and $\mathcal{F}$, define on each face or edge of V, respectively*

$$F^c(V) = \int_{S^r} F(x, (\widetilde{S, x})) \, ds + \int_{L^r} \mathcal{F}(x, (\widetilde{S, x})) \, dl. \tag{1}$$

*Remark.* We assume that $F$ and $\mathcal{F}$ depend only on the position $x$ and on the shape of $S$ at $x$.

*Remark.* We do not assume uniform boundedness of these forces in the set of all shapes.

*Remark.* A priori, $\mathcal{F}$ may depend not only on the geometry of the edge itself, but also on the limit properties of the faces whose common boundary is the considered edge.

*Regularity assumptions.* In the same way as the Cauchy's construction of the stress tensor presumes the continuity of contact forces, our construction will require the following regularity hypotheses:

 (i) On any face and any edge of an admissible surface, $F(x, (\widetilde{S, x}))$ and $\mathcal{F}(x, (\widetilde{S, x}))$ depend continuously on $x$.
 (ii) Let $E$ be a set of prescribed shapes or prescribed plane cuts. We partition $E$ into two disjoint subsets, $E^e$ and $E^r$, which contain respectively the edge and the regular shapes in $E$. We assume the equi-continuity of the families of functions $\{F(\cdot, f) : f \in E^r\}$ and $\{\mathcal{F}(\cdot, f) : f \in E^e\}$, that is, we assume that for all $\varepsilon > 0$, there exists $\eta > 0$ such that, for all $x_0$, for all $x \in B_\eta(x_0)$ (ball of radius $\eta$ and center $x_0$),

$$\forall f \in E^r, \| F(x, f) - F(x_0, f) \| < \varepsilon \quad \text{and}$$

$$\forall f \in E^e, \| \mathcal{F}(x, f) - \mathcal{F}(x_0, f) \| < \varepsilon.$$

(iii) Let $S$ be a given face of an admissible surface, let $u$ be a unit vector nowhere normal to $S$. We assume that $\mathcal{F}(x, Cut((\widetilde{S, x}), u))$ is a continuous function of the variable $x$: $\forall \varepsilon > 0, \ \forall x_0 \in S, \ \exists \eta > 0 : \ \forall x \in S$,

$$\| x - x_0 \| < \eta \Longrightarrow \| \mathcal{F}(x, Cut((\widetilde{S, x}), u) - \mathcal{F}(x_0, Cut((\widetilde{S, x_0}), u) \| < \varepsilon.$$

PROPOSITION 1. *Let S and L be a face and an edge of an admissible surface, let $u$ be a unit vector nowhere normal to S, and let B be a compact set. The function $F(x, (\widetilde{S, y}))$, $\mathcal{F}(x, (\widetilde{S, y}))$ and $\mathcal{F}(x, Cut((\widetilde{S, y}), u))$ are uniformly continuous on $B \times (B \cap S)$, $B \times (B \cap L)$ and $B \times (B \cap S)$.*

To prove the f rst part of this proposition let us denote the translated surface $S' = t_{x-y}(S)$ and the translated point $t' = t_{x-y}(t)$. It is then enough to remark that

$$\| F(x, (\widetilde{S, y})) - F(z, (\widetilde{S, t})) \| = \| F(x, (\widetilde{S', x})) - F(z, (\widetilde{S', t'})) \|$$

$$\leqslant \| F(x, (\widetilde{S', x})) - F(t', (\widetilde{S', t'})) \| + \| F(t', (\widetilde{S', t'})) - F(z, (\widetilde{S, t'})) \|$$

and to use the equi-uniform continuity (ii) and the uniform continuity (i). The same reasoning holds for $\mathcal{F}(x, (\widetilde{S, y}))$ or $\mathcal{F}(x, Cut((\widetilde{S, y}), u))$.

PROPOSITION 2. *Regularity hypotheses (i), (ii) and (iii) imply the uniform boundedness of $F(x, (\widetilde{S, x}))$ and $\mathcal{F}(x, (\widetilde{S, x}))$ on every family of admissible surfaces whose shapes are prescribed shapes or prescribed plane cuts.*

Indeed, Proposition 1 implies that, for every admissible surface $S$, $F(x, (\widetilde{S, y}))$, $\mathcal{F}(x, (\widetilde{S, y}))$ and $\mathcal{F}(x, Cut((\widetilde{S, y}), u))$ are bounded respectively on $B \times (B \cap S^r)$, $B \times (B \cap L^r)$ and $B \times (B \cap S^r)$. The proposition is proved recalling the definitio of prescribed shapes and prescribed plane cuts.

*Remark.* Proposition 2 states the relative compactness of the families $\{F(x, f): f \in E^r\}$ and $\{\mathcal{F}(x, f): f \in E^e\}$ (Corollary to the Arzelà-Ascoli theorem [17]).

*Remark.* Our hypotheses allow for any continuous dependence of surface contact forces on the curvature tensor or on any other higher order shape operator of the contact surface.

## 3. A Seeming Impossibility For Edge Forces

An important assumption for obtaining restrictions upon the dependence of contact forces on shape is the hypothesis of 'quasi-balance of contact forces' [1]. It states that there exists a positive scalar $K$ such that, for any admissible domain $V$, the following inequality holds

$$\| F^c(V) \| < K|V|. \tag{2}$$

We strengthen the preceding hypothesis, assuming the power $P_U^c$ of contact forces distribution in any $C^\infty$ velocity fi ld $U$ to be quasi-balanced. Then, for any $C^\infty$ fiel $U$, we assume the existence of a positive $K_U$ such that, for any admissible domain $V$, the following inequality holds (v.w denoting the inner product between the vectors **v** and **w**)

$$| P_U^c(V) | = \left| \int_{S^r} F(x, (\widetilde{S, x})).U(x)\,\mathrm{d}s + \int_{L^r} \mathcal{F}(x, (\widetilde{S, x})).U(x)\,\mathrm{d}l \right| < K_U|V|. \tag{3}$$

*Remark.* That this hypothesis is stronger than (2) can be verifie by considering three linearly independent constant fie ds. The dependence of $K_U$ on $U$ will be immaterial in what follows.

*Remark.* The quasi-balance of torque can be obtained by considering three independent spins.

As hypothesis (3) is stronger than (2) it will imply more stringent restrictions upon the dependence of contact force on shape. Our goal is to study its consequences for the functions $F$ and $\mathcal{F}$. We begin by considering edges whose shape is dihedral; we will then extend our results to general edges.

### 3.1. Forces on dihedral edges

THEOREM 1. *Inequality* (3) *is incompatible with nonzero contact line forces on dihedral edges: indeed* (3) *implies for any dihedral shape d,*

$$\mathcal{F}(\cdot, d) = 0. \tag{4}$$

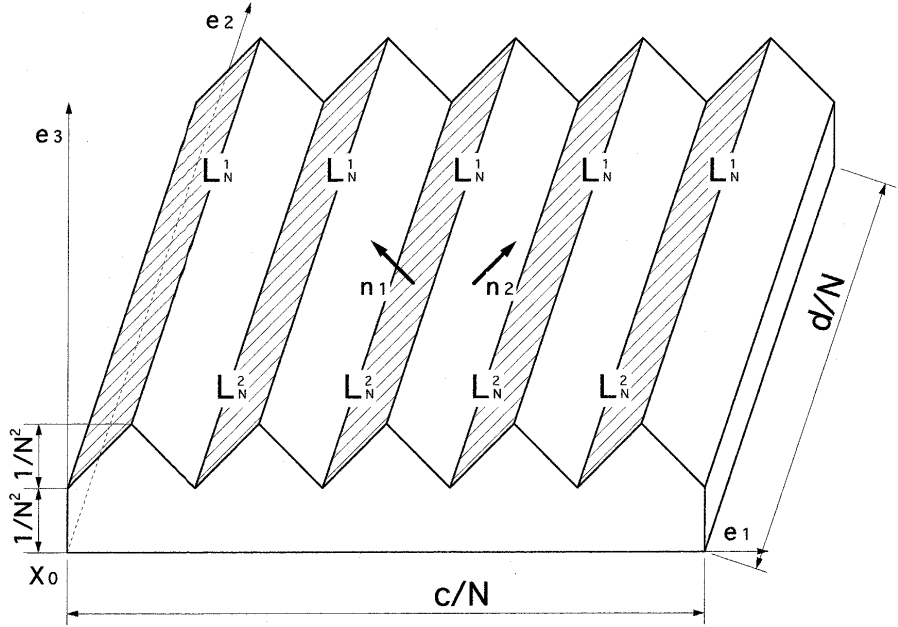

*Figure 1.* Grooved surface close to a plane

*Proof.* Let $d = (n_1, n_2, \tau)$. We use the orthogonal co-ordinate system $(x_0; e_1, e_2, e_3)$, with

$$e_2 = \tau, \qquad e_3 = \frac{n_1 + n_2}{\|n_1 + n_2\|}.$$

We consider a family of domains, parametrized by the set of integers greater than a positive $N$. The general element $V_N$ (whose boundary we denote by $S_N$) of this family is a thin slab with a grooved surface (see Figure 1). This domain is a polyhedron conceived in such a way that the set of shapes of its boundary is finite and is independent of $N$ (This set contains exactly 7 different plane shapes and 16 different dihedral shapes.) Its volume $|V_N|$ is of the same order as $N^{-4}$ when $N$ tends to infinity, the total area of its boundary $|S_N|$ is of the same order as $N^{-2}$. Let us define the following unions of edges: $L_N^1 = \{x \in S_N : \widetilde{(S, x)} = (n_1, n_2, \tau)\}$, $L_N^2 = \{x \in S_N : \widetilde{(S, x)} = (n_2, n_1, \tau)\}$ and $L_N^3 = L_N \setminus (L_N^1 \cup L_N^2)$. The total length of $L_N^3$ is of the same order as $N^{-1}$ and the total lengths of $L_N^1$ and $L_N^2$ tend to 1. Then the forces on the double array of edges $L_N^1 \cup L_N^2$ are dominant. As the shapes of $V_N$ are prescribed, contact force densities are bounded independently of $N$. Inequality (2) applied to $V_N$ implies

$$\lim_{N \longrightarrow \infty} \left\{ \int_{L_N^1} \mathcal{F}(x, (n_1, n_2, \tau)) \, dl + \int_{L_N^2} \mathcal{F}(x, (n_2, n_1, \tau)) \, dl \right\} = 0.$$

Using the mean value theorem for each component of the last equality and again the continuity of $\mathcal{F}$ with respect to $x$ we get an action-reaction principle

$$\mathcal{F}(x_0, (n_1, n_2, \tau)) = -\mathcal{F}(x_0, (n_2, n_1, \tau)). \tag{5}$$

Consider the f eld $U : x \mapsto (x.e_3) U_0$, $U_0$ being any vector. On $V_N$, $N^2 U$ is bounded independently of $N$. The same reasoning as before shows that inequality (3) implies

$$\lim_{N \to \infty} N^2 \left\{ \int_{L_N^1} \mathcal{F}(x, (n_1, n_2, \tau)).U(x) \, \mathrm{d}l + \int_{L_N^2} \mathcal{F}(x, (n_2, n_1, \tau)).U(x) \, \mathrm{d}l \right\} = 0.$$

On $L_N^1$ and $L_N^2$, $N^2 U$ does not depend either on $N$ or on $x$ as it is equal respectively to $U_0$ and $2U_0$. Then we obtain, because of arbitrariness of $U_0$,

$$\lim_{N \to \infty} \left\{ \int_{L_N^1} \mathcal{F}(x, (n_1, n_2, \tau)) \, \mathrm{d}l + 2 \int_{L_N^2} \mathcal{F}(x, (n_2, n_1, \tau)) \, \mathrm{d}l \right\} = 0.$$

Using the continuity of $\mathcal{F}$ with respect to $x$, the mean value theorem for each component of the previous equality and Equation (5) we get

$$\mathcal{F}(x_0, (n_1, n_2, \tau)) = 0. \qquad \square$$

*Remark.* This proof is not the simplest one can conceive (see the proof of Theorem 2). However, we present it here because it is suggestive: our construction shows that a limit of pairs of opposite edge forces cannot be quasi-balanced, thus illustrating the terms 'double forces' and 'double normal traction' introduced by Germain [16].

### 3.2. FORCES ON GENERAL EDGES

THEOREM 2. *Let S be an admissible surface, inequality* (3) *implies that, at every regular point $x_0$ of an edge, we have*

$$\mathcal{F}(x_0, \widetilde{(S, x_0)}) = 0 \tag{6}$$

*that is, inequality* (3) *is not compatible with the existence of contact edge forces.*

*Proof.* Let $V$ be an admissible domain whose boundary $S$ contains an edge $L$, let $x_0$ be a regular point of this edge. $S$ is tangent at the point $x_0$ to the dihedral shape $(n_1, n_2, \tau)$. In this proof we consider the case when the dihedral angle belongs to $]0, \pi[$ (The proof has to be slightly modifie if the angle is greater than $\pi$.) We use the co-ordinate system $(x_0; e_1, e_2, e_3)$ with $e_2 = \tau$,: $e_3 = n_1 + n_2/\|n_1 + n_2\|$. For any $\varepsilon > 0$, let us translate $V$ relatively to vector $\varepsilon^2 e_3$: $V' := t_{\varepsilon^2 e_3}(V)$, $S' := t_{\varepsilon^2 e_3}(S)$, $L' := t_{\varepsilon^2 e_3}(L)$ and let us defin $V^\varepsilon$ as the intersection of the domain $V'$ and of the parallelepiped $P_\varepsilon = [-c\varepsilon^2, c\varepsilon^2] \times [0, \ell\varepsilon] \times [0, 2\varepsilon^2]$ (cf. Figure 2). The dihedral angle belongs to $]0, \pi[$ and the curvatures of the faces of $S$ and of the edge $L$ are bounded. Then $c$ and $\ell$ may be chosen in such a way that, for $\varepsilon$ small enough:

(i) $L'$ meets $\partial P_\varepsilon$ on the surfaces $\{x.e_2 = 0\}$ and $\{x.e_2 = \ell \varepsilon\}$, then at every point $x$ on $L' \cap P_\varepsilon$, $x.e_3 > 0$ holds,
(ii) $S'$ meets $\partial P_\varepsilon$ on surfaces $\{x.e_2 = 0\}$, $\{x.e_2 = \ell \varepsilon\}$ and on the surface $\{x.e_3 = 0\}$.

These properties are represented in Figure 2. We denote by $S_\varepsilon$ the boundary of $V_\varepsilon$ and by $L_\varepsilon$ the upper edge of $V_\varepsilon$ ($L_\varepsilon = L' \cap P_\varepsilon$). The shapes of the boundary $S_\varepsilon$ of $V_\varepsilon$ are either prescribed shapes or prescribed plane cuts. Then the surface force density $F$ and the line force density $\mathcal{F}$ are bounded independently of $\varepsilon$.

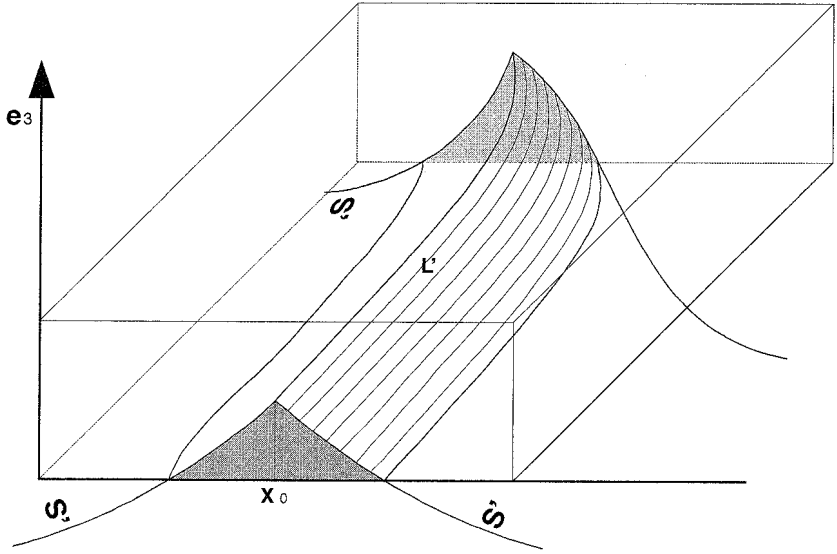

*Figure 2*. The vicinity of an edge

Let us consider the vector field $U : x \mapsto (x.e_3) U_0$, $U_0$ being any vector. The geometry of $V_\varepsilon$ assures that $\varepsilon^{-2} U$ is bounded independently of $\varepsilon$. On the other hand, as $U$ vanishes on the plane $(x_0, e_1, e_2)$, we do not have to consider the forces exerted on the edges which are included in this plane. Considering the measure of each face and edge, we get from inequality (3)

$$\lim_{\varepsilon \to 0} \varepsilon^{-3} \int_{L_\varepsilon} \mathcal{F}(x, (\widetilde{S_\varepsilon, x})).U(x) \, \mathrm{d}l = 0.$$

The length of $L_\varepsilon$ is $\varepsilon \ell$ to within higher order terms. On the other hand, there exists a positive scalar $k$ (depending on the curvature of the edge at $x_0$ and on $\ell$ but independent of $\varepsilon$) such that

$$\lim_{\varepsilon \to 0} \varepsilon^{-3} \int_{L_\varepsilon} (x.e_3) \, \mathrm{d}l = k.$$

Let $\delta > 0$, the geometry of the domain and Proposition 1 – consequence of the regularity assumptions – imply that, for $\varepsilon$ small enough,

$$\forall x \in L_\varepsilon, \qquad \|\mathcal{F}(x, (\widetilde{S_\varepsilon, x})) - \mathcal{F}(x_0, (\widetilde{S, x_0}))\| < \delta.$$

Then

$$\left| k \, \mathcal{F}(x_0, (\widetilde{S, x_0})).U_0 - \lim_{\varepsilon \to 0} \varepsilon^{-3} \int_{L_\varepsilon} \mathcal{F}(x, (\widetilde{S_\varepsilon, x})).U(x) \, \mathrm{d}l \right| < k \, \delta \, \|U_0\|.$$

This result holds for any $\delta$ and for any $U_0$, so that

$$\mathcal{F}(x_0, (\widetilde{S, x_0})) = 0. \qquad \square$$

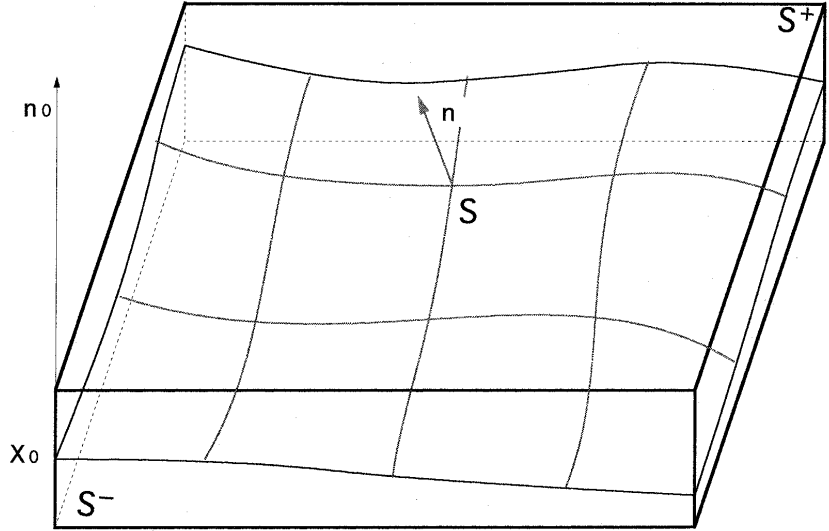

*Figure 3*. The domain used in the proof of Noll Theorem

### 3.3. NOLL THEOREM

Theorem 2 states that there are no contact edge forces. Then Noll theorem can be proved. Here we give a proof valid in the case when the surface contact force density is not bounded independently of the considered contact surface.

THEOREM 3. *When no edge forces are present, and under the regularity assumptions listed in Section* 2, *for all regular shapes f tangent to the plane shape n we have*

$$F(x, f) = F(x, n). \tag{7}$$

*Proof.* Let $S$ be the boundary of an admissible domain $V$ and let $x_0$ be a regular point of $S^r$. We call $n_0$ the normal to $S$ at $x_0$. We use the co-ordinate system $(x_0, e_1, e_2, e_3)$ (with $e_3 = n_0$). Let us consider the family of parallelepipeds $C_\varepsilon = [0, \varepsilon] \times [0, \varepsilon] \times [-c\varepsilon^2, +c\varepsilon^2]$. We defin $V_\varepsilon = V \cap C_\varepsilon$ and $S_\varepsilon = S \cap C_\varepsilon$. As the curvature of $S$ is bounded in a neighbourhood of $x_0$, a positive scalar $c$ can be found such that, for $\varepsilon$ suff ciently small, $S_\varepsilon$ does not intersect the face $(S^+ = \{x_3 = c\varepsilon^2\})$ or the face $(S^- = \{x_3 = -c\varepsilon^2\})$ of $C_\varepsilon$ (cf. Figure 3). The shapes of the boundary of $V_\varepsilon$ are either prescribed shapes or prescribed plane cuts.

Inequality (2) when applied to $C_\varepsilon$ implies

$$\lim_{\varepsilon \to 0} \varepsilon^{-2} \int_{S^+} F(x, n_0)\, \mathrm{d}s + \lim_{\varepsilon \to 0} \varepsilon^{-2} \int_{S^-} F(x, -n_0)\, \mathrm{d}s = 0,$$

which leads to

$$F(x, n_0) = -F(x, -n_0). \tag{8}$$

Inequality (2) when applied to $V_\varepsilon$ implies

$$\lim_{\varepsilon \to 0} \varepsilon^{-2} \int_{S_\varepsilon} F(x, \widetilde{(S, x)})\, \mathrm{d}s + \lim_{\varepsilon \to 0} \varepsilon^{-2} \int_{S^-} F(x, -n_0)\, \mathrm{d}s = 0,$$

which leads to

$$F(x_0, (\widetilde{S, x_0})) + F(x_0, -n_0) = 0.$$

Equation (7) is then obtained when recalling (8). □

*Remark.* In the previous proof we just had to modify the argument used in [3] by using a cylinder whose basis is a square instead of a circle. The important difference is that we use only prescribed shapes (in the sense define in Section 2). In the proof of Noll Theorem as in the proof of Cauchy Theorem we do not use assumption (3) but only assumption (2).

3.4. CAUCHY THEOREM

Theorem 2 states that there are no contact edge forces. Then the Cauchy's construction of stress tensor (refer to [3], [5]) is valid.

THEOREM 4. *When no edge force is present, there exists a continuous two-tensor fie d $T$ such that, for any plane shape $n$,*

$$F(x, n) = T(x).n. \tag{9}$$

## 4. Contact Force Distribution of Order One with respect to Normal Derivative

4.1. ON CONTACT FORCE DISTRIBUTIONS OF ORDER ONE WITH RESPECT TO NORMAL DERIVATIVE, EXAMPLES, HYPOTHESES

The previous section has shown the impossibility of introducing contact edge forces under the assumption (3). The idea we will develop now is the following: the power of contact force distribution is actually quasi-balanced but the expression for this power used in inequality (3) is naive; the contact force distribution must be endowed with a more complex structure. We assume that $F(x, (\widetilde{S, x}))$ and $\mathcal{F}(x, (\widetilde{S, x}))$ represent only the zero order part of a more general distribution whose support is $S$. In this paper we add to the considered distribution another one whose order is one with respect to normal derivative of test functions. More precisely we add a distribution whose value at a vector fie d $U$ is $\int_{S^r} G(x, (\widetilde{S, x})).\partial U/\partial n(x)\,\mathrm{d}s$.

A decomposition Theorem [14] states that every distribution $D$ whose support is a smooth surface (i.e. a $C^\infty$ submanifold of codimension one) can be decomposed as follows

$$DU = \sum_{i=0}^{N} D_i \left(\frac{\partial^i U}{\partial n^i}\right), \tag{10}$$

where $D_i$ are uniquely determined surface zero-order distributions. We limit ourselves to distributions of order one ($N = 1$) and we postpone the discussion on the influenc of higher order distributions until Section 5.

Let us remark that, from a physical point of view, the introduction of a contact force distribution '1-normal' can be interpreted as follows: in the balance of energy an additional term is needed which does not appear in the balance of forces. An alternative approach satisfying this need is due to Dunn and Serrin [2], who introduced directly a supplementary fl x of energy ('interstitial working'). Our presentation, based on the concept of distributions, has the following features: (i) it does not assume a priori that the extra energetic term is a f ux, (ii) it shows the mechanical nature of this term, its linear dependence on the velocity fi ld being a basic assumption, (iii) it naturally yields general and physically meaningful boundary conditions [11].

Most mechanicians will not be surprised by the introduction of contact forces distributions (in the mathematical sense), as contact couples are needed already in the standard theories of beams and shells. Another example of '1-normal' contact forces distribution found in the literature (this time for 3-D continua) is given by couple stresses introduced by Cosserat [13]. The microscopic meaning of a distribution of order greater than zero can be understood by considering the asymptotic limit of non-local short range interactions.

### 4.2. A TRIVIAL EXAMPLE OF NON-LOCAL SHORT RANGE FORCES CONVERGING TO A '1-NORMAL' DISTRIBUTION

Using the Cartesian co-ordinates $(x_1, x_2, x_3)$, the domain $V = \{x: x_1 < 0\}$ is divided from the external world by the plane $S = \{x: x_1 = 0\}$. Assume that the external forces exerted on $V$ have short range $\varepsilon \ll 1$ (compared with some other characteristic length) and that these forces are represented by the volume density $f_\varepsilon(x) = f_0\, \varepsilon^{-\gamma} \varphi(\varepsilon^{-1} x_1)$ where $\varphi$ is a function whose support is a compact set included in $]-\infty, 0[$ and $f_0$ is a given vector. If $\gamma = 1$ and $\varphi$ is a non negative function whose integral is equal to 1, the distribution tends, as $\varepsilon$ tends to 0, to the vector measure on $S$ whose surface density is $f_0$: this is the classical case of surface force density. However, if $\gamma = 2$ and $\varphi$ is the derivative of a non negative function whose integral is equal to 1, the distribution tends to a '1-normal' distribution $D$ such that $D(U) = \int_S f_0.\partial U/\partial x_1\, \mathrm{d}s$. This force distribution

 (i) is localized on $S$,
 (ii) has no influenc upon balance of forces,
 (iii) supplies energy in presence of velocity field in $V$, even when these f elds vanish on $S$.

### 4.3. 'QUASI-BALANCE' OF CONTACT FORCES DISTRIBUTION

Then we relax the hypothesis formulated in Section 3, in order to allow the presence of a '1-normal' distribution on the boundary $S$ of an admissible domain $V$. Contact forces interaction is described, in addition to $F(x, (\widetilde{S, x}))$ and $\mathcal{F}(x, (\widetilde{S, x}))$ introduced in Section 2, by a surface density $G(x, (\widetilde{S, x}))$ on the regular part of the contact surface. The power $P_U^c(V)$ of this interaction in a velocity fi ld $U$ is given by the following expression

$$\int_{S^r} G(x, (\widetilde{S, x})).\frac{\partial U}{\partial n}(x)\, \mathrm{d}s + \int_{S^r} F(x, (\widetilde{S, x})).U(x)\, \mathrm{d}s + \int_{L^r} \mathcal{F}(x, (\widetilde{S, x})).U(x)\, \mathrm{d}l.$$

We assume the same regularity properties for $G$ as we did for $F$ in Section 2.

The assumption of 'quasi-balance' of contact forces distribution is formulated as follows: for all $C^\infty$ fiel $U$, there exists a scalar $K_U$ such that, for any admissible domain $V$, we have

$$\left| \int_{S^r} G(x, (\widetilde{S, x})).\frac{\partial U}{\partial n}(x)\, \mathrm{d}s + \int_{S^r} F(x, (\widetilde{S, x})).U(x)\, \mathrm{d}s \right.$$
$$\left. + \int_{L^r} \mathcal{F}(x, (\widetilde{S, x})).U(x)\, \mathrm{d}l \right| < K_U |V|. \tag{11}$$

*Remark.* This assumption is less stringent on $\mathcal{F}$ than the corresponding hypothesis (3). It will imply less stringent restrictions upon edge contact forces.

*Remark.* The fact that (11) is stronger than the quasi-balance of forces (2) is still true (it can be verified by considering three linearly independent constant fields $U$). Again, we do not need any assumption on the behaviour of $K_U$ with respect to $U$.

*Remark.* Note that, in each proof in this paper, we pass to the limit in inequality (11) with a fixed field $U$. For this reason we do not need any assumption on the behaviour of $K_U$ with respect to $U$.

4.4. DEPENDENCE OF NORMAL DISTRIBUTION ON THE SHAPE OF THE CONTACT SURFACE. A THEOREM ANALOGOUS TO NOLL THEOREM

We first prove an action-reaction principle which, at this stage, concerns only plane shapes.

THEOREM 5. *At every point $x$ and for all plane shape $n$ we have*

$$G(x, n) = G(x, -n). \tag{12}$$

*Proof.* Using the co-ordinate system $(x_0, e_1, e_2, e_3)$ (with $e_3 = n$), let us consider the domain $C_\varepsilon = [0, \varepsilon] \times [0, \varepsilon] \times [0, \varepsilon^2]$ and the vector field $U: x \mapsto x.n\, U_0$ where $U_0$ is a given vector. The shapes of $\partial C_\varepsilon$ are prescribed shapes. Moreover $\varepsilon^{-2} U$ is bounded on $C_\varepsilon$ independently of $\varepsilon$. Considering the area or length of each face or edge, the inequality (11) applied to $C_\varepsilon$ implies

$$\lim_{\varepsilon \to 0} \varepsilon^{-2} \int_{S^+} G(x, n).U_0 \, ds + \lim_{\varepsilon \to 0} \varepsilon^{-2} \int_{S^-} G(x, -n).(-U_0) \, ds = 0,$$

where $S^+$ and $S^-$ denote the upper and lower faces of $P_\varepsilon$ ($S^+ = \{x: x.e_3 = +\varepsilon^2\}$, $S^- = \{x: x.e_3 = 0\}$). The continuity properties and the arbitrariness of $U_0$ imply

$$G(x_0, n) = G(x_0, -n). \qquad \square$$

We are now able to prove a theorem for '1-normal' distributions analogous to the theorem of Noll [3], [18].

THEOREM 6. *At every point $x$ and for every regular shape $f$ tangent to the plane shape $n$ we have*

$$G(x, f) = G(x, n); \tag{13}$$

*that is, G depends on the shape of the contact surface only through its normal.*

*Proof.* The proof is close to that we have given for the Noll Theorem. At a regular point $x_0$ of the boundary $S$ of an admissible domain, we consider the family of domains $V_\varepsilon$ described in the proof of the Noll Theorem or in Figure 3. We consider the vector field $U: x \mapsto x.n_0\, U_0$ where $n_0$ denotes the normal to $S$ at $x_0$ and $U_0$ is any vector. The shapes of $\partial V_\varepsilon$ are either prescribed shapes or prescribed plane cuts. Moreover $\varepsilon^{-2}U$ is bounded in $V_\varepsilon$ independently of $\varepsilon$. The inequality (11) applied to $V_\varepsilon$ leads to

$$\lim_{\varepsilon \to 0} \varepsilon^{-2} \int_{S \cap C_\varepsilon} G(x, \widetilde{(S, x)}).U_0(n.n_0)\, \mathrm{d}s + \lim_{\varepsilon \to 0} \varepsilon^{-2} \int_{S^-} G(x, -n_0).(-U_0)\, \mathrm{d}s = 0.$$

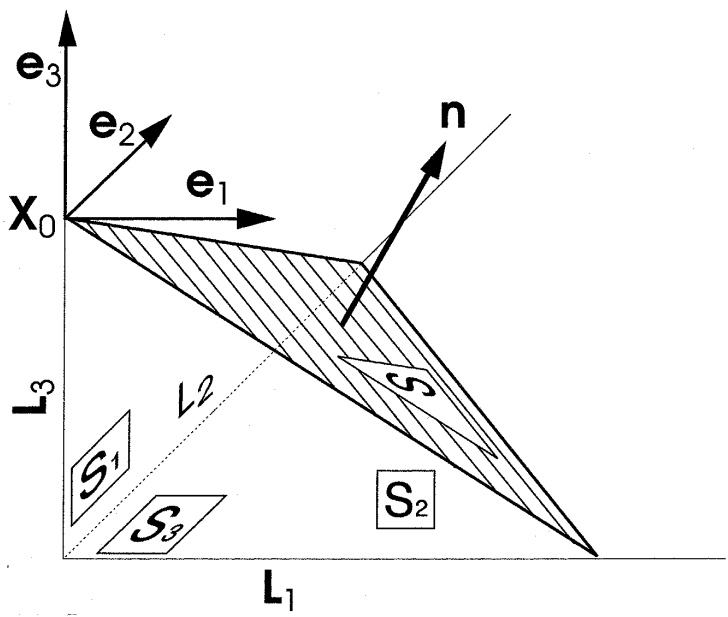

*Figure 4.* Cauchy's tetrahedron

As $n.n_0$ is a continuous function with respect to $x$ on $S$, the regularity properties of $G$ and arbitrariness of $U_0$ imply

$$G(x_0, \widetilde{(S, x_0)}) = G(x_0, -n_0) = G(x_0, n_0). \qquad \square$$

### 4.5. A CAUCHY THEOREM FOR HYPERSTRESS

THEOREM 7. *There exists a continuous three-tensor field $C$ such that, at any point $x_0$ and for any plane shape $n$,*

$$G(x_0, n) = (C(x_0).n).n. \tag{14}$$

*Proof.* We follow the tetrahedron construction due to Cauchy (see Figure 4). In an orthonormal co-ordinate system $(x_0, e_1, e_2, e_3)$, we define the tetrahedron $V$ whose faces $S, S_1, S_2$ and $S_3$ are respectively normal to $n, -e_1, -e_2$ and $-e_3$ and whose height (perpendicular to $S$) is $h$. In our construction the origin of the co-ordinate system belongs to the face $S$. We denote respectively by $f_1 = (-e_2, -e_3, e_1)$, $f_2 = (-e_3, -e_1, e_2)$ and $f_3 = (-e_1, -e_2, e_3)$ the shapes of the edges $L_1, L_2,$ and $L_3$. Let $V_\varepsilon$ be the image of $V$ under an homothetic transformation of ratio $\varepsilon$, we denote by $S_\varepsilon, S_{i\varepsilon}$ and $L_{i\varepsilon}$ the faces and edges images of $S, S_i$ and $L_i$ ($i = 1, 2, 3$) under this transformation. We consider the field $U : x \mapsto (x.n)U_0$, where $U_0$ is any vector. As this field vanishes on $S_\varepsilon$, the inequality (11) applied to the domain $V_\varepsilon$ implies

$$\left| \sum_{i=1}^{3} \left\{ \int_{S_{i\varepsilon}} (x.n)\, F(x, -e_i).U_0 \,\mathrm{d}s \right\} + \sum_{i=1}^{3} \left\{ \int_{L_{i\varepsilon}} (x.n)\, \mathcal{F}(x, f_i).U_0 \,\mathrm{d}l \right\} \right.$$

$$\left. + \sum_{i=1}^{3} \left\{ \int_{S_{i\varepsilon}} (-e_i.n)\, G(x, -e_i).U_0 \,\mathrm{d}s \right\} + \int_{S_\varepsilon} G(x, n).U_0 \,\mathrm{d}s \right| < K\varepsilon^3.$$

Let us multiply this inequality by $\varepsilon^{-2}$ and, changing variables in the integrals in order to transform them into integrals on the boundary of $V$, we obtain

$$\left| \varepsilon \sum_{i=1}^{3} \left\{ \int_{S_i} (x.n)\, F(\varepsilon x, -e_i).U_0 \,\mathrm{d}s \right\} + \sum_{i=1}^{3} \left\{ \int_{L_i} (x.n)\, \mathcal{F}(\varepsilon x, f_i).U_0 \,\mathrm{d}l \right\} \right.$$

$$\left. + \sum_{i=1}^{3} \left\{ \int_{S_i} (-e_i.n)\, G(\varepsilon x, -e_i).U_0 \,\mathrm{d}s \right\} + \int_{S} G(\varepsilon x, n).U_0 \,\mathrm{d}s \right| < K\varepsilon.$$

As $F(., f)$ and $\mathcal{F}(., f)$ are continuous, evaluating the limit as $\varepsilon$ tends to 0, we get

$$\sum_{i=1}^{3} \left\{ \mathcal{F}(x_0, f_i).U_0 \int_{L_i} (x.n) \,\mathrm{d}l \right\} + \sum_{i=1}^{3} \{|S_i|(-e_i.n)G(x_0, -e_i).U_0\}$$

$$+|S|G(x_0, n).U_0 = 0.$$

This being valid for any $U_0$, we obtain

$$2|S|G(x_0, n) = \sum_{i=1}^{3} \{\mathcal{F}(x_0, f_i)(n.e_i)|L_i|^2\} + 2\sum_{i=1}^{3} \{G(x_0, -e_i)|S_i|(n.e_i)\}.$$

Using the geometrical relations

$$h = |L_1|(n.e_1) = |L_2|(n.e_2) = |L_3|(n.e_3),$$

$$2|S|h = 2|S_1||L_1| = 2|S_2||L_2| = 2|S_3||L_3| = |L_1||L_2||L_3|$$

and Theorem 5, we get

$$G(x_0, n) = \mathcal{F}(x_0, f_1)(n.e_2)(n.e_3) + \mathcal{F}(x_0, f_2)(n.e_3)(n.e_1)$$

$$+\mathcal{F}(x_0, f_3)(n.e_1)(n.e_2) + \sum_{i=1}^{3} G(x_0, e_i)(n.e_i)^2. \tag{15}$$

Thus we are led to define a three-tensor field $C$ such that $G(x_0, n) = (C(x_0).n).n$. This tensor is not uniquely determined, as only its right-side products by symmetric two tensors are determined. We may impose its right side symmetry, setting

$$C(x) = \tfrac{1}{2}\mathcal{F}(x, f_1) \otimes (e_2 \otimes e_3 + e_3 \otimes e_2) + \tfrac{1}{2}\mathcal{F}(x, f_2) \otimes (e_3 \otimes e_1 + e_1 \otimes e_3)$$

$$+\tfrac{1}{2}\mathcal{F}(x, f_3) \otimes (e_1 \otimes e_2 + e_2 \otimes e_1) + \sum_{i=1}^{3} \{G(x, e_i) \otimes e_i \otimes e_i\}, \tag{16}$$

or its left side symmetry, setting

$$C(x) = \tfrac{1}{2}\mathcal{F}(x, f_1) \otimes (e_2 \otimes e_3 + e_3 \otimes e_2) - \tfrac{1}{2}(e_2 \otimes e_3 + e_3 \otimes e_2) \otimes \mathcal{F}(x, f_1)$$

$$+ \tfrac{1}{2}(e_2 \otimes \mathcal{F}(x, f_1) \otimes e_3 + e_3 \otimes \mathcal{F}(x, f_1) \otimes e_2) + \tfrac{1}{2}\mathcal{F}(x, f_2)$$

$$\otimes (e_3 \otimes e_1 + e_1 \otimes e_3)$$

$$- \tfrac{1}{2}(e_3 \otimes e_1 + e_1 \otimes e_3) \otimes \mathcal{F}(x, f_2) + \tfrac{1}{2}(e_3 \otimes \mathcal{F}(x, f_2)$$

$$\otimes e_1 + e_1 \otimes \mathcal{F}(x, f_2) \otimes e_3)$$

$$+ \tfrac{1}{2}\mathcal{F}(x, f_3) \otimes (e_1 \otimes e_2 + e_2 \otimes e_1)$$

$$- \tfrac{1}{2}(e_1 \otimes e_2 + e_2 \otimes e_1) \otimes \mathcal{F}(x, f_3)$$

$$+ \tfrac{1}{2}(e_1 \otimes \mathcal{F}(x, f_3) \otimes e_2 + e_2$$

$$\otimes \mathcal{F}(x, f_3) \otimes e_1)$$

$$+ \sum_{i=1}^{3} \{G(x, e_i) \otimes e_i \otimes e_i - G(x, e_i)$$

$$\otimes e_i \otimes e_i + e_i \otimes G(x, e_i) \otimes e_i\}. \tag{17}$$

*Remark 1.* The tensor $C$ will be called the hyperstress tensor.

*Remark 2.* Imposing the left side symmetry of $C$ seems complicated [19] and artificial but Theorem 9 will show the advantage of such a definition.

*Remark 3.* Equation (16) reduces to $G(x_0, n) = \Sigma_{i=1}^{3} G(x_0, e_i)(n.e_i)^2$ when $\mathcal{F}$ is vanishing. As this expression is valid for every orthonormal vector basis, then $G$ does not depend on $n$: this is the only case in which the 1-normal distribution can be nonzero with vanishing edge forces.

4.6. REPRESENTATION THEOREMS FOR CONTACT FORCES. CAUCHY STRESS TENSOR

In this section we add the following assumption:

*Hypothesis of $C^1$ regularity.* For every given regular shape $f$ and every dihedral shape $d$, $G(\cdot, f)$ and $\mathcal{F}(\cdot, d)$ are $C^1$ functions. In this way $C(\cdot)$ is a field of class $C^1$.

We now can show how our hypothesis (11), which implies hypothesis (2) put forward by Noll and Virga, allows us to prove the assumption III on page 21 of [1], and to show that the example treated in Section 9 always in [1] actually exhausts all possible cases. Indeed the following two Theorems give the general representation of edge and surface contact forces.

THEOREM 8. *Let $S$ be the boundary of an admissible domain $V$. Let $x$ be a regular point of an edge of $S$. Let $(n_1, n_2, \tau)$ be the tangent dihedral shape to $S$ at $x$. Then the edge force density at $x$ depends only on $(n_1, n_2, \tau)$ and is represented in terms of the second order stress tensor by*

$$\mathcal{F}(x, \widetilde{(S, x)}) = (C(x).n_1).\nu_1 + (C(x).n_2).\nu_2, \tag{18}$$

where we denote $\nu_1 = \tau \times n_1$ and $\nu_2 = -\tau \times n_2$.

*Remark.* The arbitrariness in $C$ has no influence on the representation formula (19) as $n_1 \otimes \nu_1 + n_2 \otimes \nu_2$ is a symmetric tensor.

THEOREM 9. *It exists a continuous second order tensor field $T$ such that*

$$F(x, (\widetilde{S, x})) = T(x).n - \nabla^s.(((C(x).n).\Pi), \tag{19}$$

*where $\Pi$ denotes the projector on the tangent plane to the surface ($\Pi = \mathrm{Id} - n \otimes n$) and $\nabla^s\cdot$ the surface divergence.*

*Remark.* The arbitrariness in $C$ has an influence on $T$. With the choice (18), the tensor $T$ is symmetric.

*Proof of Theorems 8 and 9.* Because of Theorem 7, inequality (11) may be written

$$\left| \int_{L^r} (\mathcal{F}(x, (\widetilde{S, x}))).U(x)\,\mathrm{d}l + \int_{S^r} F(x, (\widetilde{S, x})).U(x)\,\mathrm{d}s \right.$$
$$\left. + \int_{S^r} ((C(x).n).n).\frac{\partial U}{\partial n}(x)\,\mathrm{d}s \right| < K_U |V|.$$

Due to the equality

$$\left| \int_V \mathrm{div}(\nabla U : C)\,\mathrm{d}v \right| = \left| \int_{S^r} \nabla U : (C.n)\,\mathrm{d}s \right|, \tag{20}$$

the quantity

$$\int_{S^r} \nabla U : (C.n)\,\mathrm{d}s$$

is quasi-balanced. Then there exists a scalar $K'_U$ such that

$$\left| \int_{L^r} \mathcal{F}(x, (\widetilde{S, x})).U(x)\,\mathrm{d}l + \int_{S^r} F(x, (\widetilde{S, x})).U(x)\,\mathrm{d}s \right.$$
$$\left. + \int_{S^r} \left\{ ((C(x).n).n).\frac{\partial U}{\partial n}(x) - \nabla U(x) : (C.n) \right\}\,\mathrm{d}s \right| < K'_U |V|.$$

Using now the decomposition $\nabla U : (C.n) = \partial U/\partial n.((C.n).n) + \nabla^s U : ((C.n).\Pi)$ where $\nabla^s$ denotes the surface gradient on $S$, we get

$$\left| \int_{L^r} \mathcal{F}(x, (\widetilde{S, x})).U(x)\,\mathrm{d}l + \int_{S^r} F(x, (\widetilde{S, x})).U(x)\,\mathrm{d}s \right.$$
$$\left. + \int_{S^r} \nabla^s U(x) : ((C(x).n).\Pi)\,\mathrm{d}s \right| < K'_U |V|.$$

We apply the divergence theorem on every face of $S$, so obtaining

$$\left| \int_{L^r} \mathcal{F}'(x, (\widetilde{S, x})).U(x)\,\mathrm{d}l + \int_{S^r} F'(x, (\widetilde{S, x})).U(x)\,\mathrm{d}s \right| < K'_U |V|, \tag{21}$$

where we used the following definitions

$$\mathcal{F}'(x, (\widetilde{S, x})) := \mathcal{F}(x, (\widetilde{S, x})) - (C(x).n_1).\nu_1 - (C(x).n_2).\nu_2,$$

$$F'(x, (\widetilde{S, x})) := F(x, (\widetilde{S, x})) + \nabla^s.((C(x).n).\Pi).$$

Because of the regularity hypotheses added in this section, $F'$ and $\mathcal{F}'$ verify the assumptions of Section 3. Because of inequality (22), we may apply to them our results of Section 3. Theorem 2 implies that $\mathcal{F}'(x, (\widetilde{S, x}))$ is vanishing and Theorem 4 establishes the existence of a continuous second order tensor field $T(\cdot)$ such that

$$F'(x, (\widetilde{S, x})) = T(x).n. \qquad \square$$

## 5. Conclusions

The most important concepts introduced in this paper are:

 (i) the concept of quasi-balanced power of contact force distribution and
(ii) that of prescribed shapes.

They allowed us to develop a system of axioms 'à la Cauchy' for continua in which edge contact forces are present. In these media, stress states are described by two tensor fields, one ($T$) of order two and one ($C$) of order three. One may easily verify that the resultant contact force exerted on an admissible domain is only due to the tensor $T$. Then, in the local form of balance of forces, only the divergence of $T$ appears [11]. In this sense $T$ plays the role of the Cauchy stress tensor. On the contrary, $T$ is not able to describe surface contact forces locally, even when the contact surface is plane. For this description and for the description of contact edge forces the tensor $C$ is needed. This class of continua has already been described by Germain [15] who called them second gradient media. His starting assumption concerns the form of internal power and the form of the power of contact forces is obtained via the D'Alembert principle of virtual powers. Our quasi-balance assumption of contact force distributions is a weak form of this principle.

It is easy to verify that the density of internal power is not equal to $T : \nabla U$ but that a supplementary term equal to $\text{div}(\nabla U : C)$ appears [11]. This term, due to a flux $q = \nabla U : C$, can be identified with the interstitial working introduced by Dunn and Serrin. Our results show that the interstitial work flux through a contact surface can be interpreted as the sum of three terms:

  (i) the power of edge contact forces,
 (ii) the power of a part of surface forces (the part depending on $C$),
(iii) the power expended by the normal distribution.

We thus obtain a mechanical interpretation of interstitial working. Remark 3 of Section 4.3 makes explicit the possible expressions for interstitial work flux when edge forces are vanishing.

We obtained our results under the assumption that the normal contact force distribution was of the order of one. As we did not give any physical justification for this hypothesis, one could wonder if the representation theorems we have found for surface and edge forces are simply consequences of this assumption. This is not the case. Indeed our theorems are true when the contact force distribution is of higher order with respect to normal derivative: i.e. if in inequality (11) we substitute the expression of $P_U^c(V)$ by

$$\int_{S^r} F(x,(\widetilde{S,x})).U(x)\,\mathrm{d}s + \int_{L^r} \mathcal{F}(x,(\widetilde{S,x})).U(x)\,\mathrm{d}l + \sum_{i=1}^{N}\int_{S^r} G_i(x,(\widetilde{S,x})).\frac{\partial^i U}{\partial n^i}(x)\,\mathrm{d}s$$

the functions $G_i$ for $i > 1$ do not play any role in our demonstrations as we only use affin vector fie ds $U$.

We also assumed that there are no transverse distributions on the edges. One could wonder how our results are influence by this assumption. If we add to $P_U^c(V)$ the term $\Sigma_{i=1}^{N}\int_{L^r} H_i(x,(\widetilde{S,x}))\nabla_T^i U(x)\,\mathrm{d}l$, where $\nabla_T$ denotes normal gradient on the edge whose tangent is $\tau$ ($\nabla_T = (\mathrm{Id} - \tau \otimes \tau).\nabla$), then also the functions $H_i$ for $i > 1$ do not play any role for the same reason. We did not yet studied the influenc of a non vanishing $H_1$ on our results.

The remaining open problems are the study of the influenc of a non-vanishing distribution of the order of one on the edges ($H_1$) and the study of the influenc of forces concentrated at vertices (intersections of edges). Indeed the results expressed as Theorems 6 and 7 are most likely due to the assumption of absence of forces concentrated at vertices in the same way as the Noll Theorem and Cauchy representation Theorem are valid only when edge forces are vanishing. A general classifica ion of quasi-balanced distributions seems an interesting development of this paper.

## Acknowledgements

We thank C. Gavarini, Director of the Dipartimento di Ingegneria Strutturale e Geotecnica, dell'Università di Roma 'La Sapienza', for having assured a visiting grant to one of the authors, and G. Maugin and W. Kosinski for their friendly criticism. We are greatly indebted to A. Di Carlo: his valuable and careful criticism was very useful in improving this paper.